\documentclass[showpacs,twocolumn]{revtex4}
\usepackage{epsfig}
\def\beq{\begin{equation}}
\def\enq{\end{equation}}
\def\beqa{\begin{eqnarray}}
\def\enqa{\end{eqnarray}}

\def\MeV{\nobreak\,\mbox{MeV}}
\def\GeV{\nobreak\,\mbox{GeV}}

\def\qq{\lag\bar{q}q\rag}

\def\sss{\lag\bar{s}s\rag}
\def\qqs{\lag\bar{s}s\rag}
\def\mix{\lag\bar{q} G q\rag}
\def\mixs{\lag\bar{s} G s\rag}
\def\Gd{\lag g_s^2G^2\rag}
\def\G3{\lag g_s^3G^3\rag}

\def\rh{\rho}

\def\al{\alpha}
\def\be{\beta}
\def\alma{\alpha_{max}}
\def\almi{\alpha_{min}}
\def\bemi{\beta_{min}}

\def\lb{\label}
\def\nn{\nonumber}

\def\xsla{x\kern-.5em\slash}
\def\psla{p\kern-.5em\slash}
\newcommand{\rag}{\rangle}
\newcommand{\lag}{\langle}

\begin{document}

\title{\sc Exotic $1^{--}$ States in QCD Sum Rules}
\author{Raphael M. Albuquerque}
\email{rma@if.usp.br}
\affiliation{Instituto de F\'{\i}sica, Universidade de S\~{a}o Paulo,
C.P. 66318, 05315-970 S\~{a}o Paulo, SP, Brazil}
\author{Marina Nielsen}
\email{mnielsen@if.usp.br}
\affiliation{Instituto de F\'{\i}sica, Universidade de S\~{a}o Paulo,
C.P. 66318, 05315-970 S\~{a}o Paulo, SP, Brazil}
\author{R\^omulo Rodrigues da Silva}
\email{romulo@df.ufcg.edu.br}
\affiliation{
UAF, Universidade Federal de Campina Grande\\
58.051-970 Campina Grande, PB, Brazil}
\begin{abstract}
Using the QCD  sum rules we test if the charmonium-like structure 
$Y(4260)$, observed in the $J/\psi\pi\pi$ invariant mass spectrum, can 
be described as a exotic state, with a $J/\psi~f_0(980)$ molecular current 
with $J^{PC}=1^{--}$. By exotic we mean a more complex structure than the 
simple quark-antiquark state and not exotic $J^{PC}$ quantum numbers.
We consider the contributions of condensates 
up to dimension six and we work at leading order in $\alpha_s$.
We keep terms which are linear in the strange quark mass $m_s$. 
The mass obtained for such state is $m_{Y}=(4.67\pm 0.09)$ 
GeV, when the vector and scalar mesons are in color singlet configurations. 
We conclude that the proposed current can better describe the $Y(4660)$ state
that could be interpreted as a $\Psi(2S)~f_0(980)$ molecular state.
We also use different $J^{PC}=1^{--}$ currents to study the recently observed 
$Y_b(10890)$ state. Our findings indicate that the $Y_b(10890)$ can be well 
described by a scalar-vector tetraquark current. 
\end{abstract}

\maketitle

\section{Introduction}
Most of the states recently observed at the $B$ factories and the Tevatron,
the $X,~Y$ and $Z$ states, do not fit the quarkonia interpretation. Their
production mechanism, masses, decay widths, spin-parity assignments and decay modes
have been discussed in some reviews 
\cite{Brambilla:2010cs,Nielsen:2009uh,Olsen:2009gi}. Particularly interesting
are the $1^{--}$ states, observed in $e^+e^-$ annihilation. The first state in the 
$1^{--}$ family discovered in the $e^+e^-$ annihilation through initial state 
radiation was the $Y(4260)$ \cite{babar1}. Repeating the same kind of analysis 
leading to the observation of the $Y(4260)$ state, in the channel $e^+e^-\to
\gamma_{ISR}\Psi(2S)\pi^+\pi^-$, BaBar \cite{babar4} has identified another broad 
peak at a mass around 4.32 GeV, which was confirmed by Belle \cite{belle4}. 
Belle found that the $\psi^\prime\pi^+\pi^-$ enhancement observed by BaBar was, in 
fact, produced by two distinct peaks the $Y(4360)$ and the $Y(4660)$. In the
bottom sector, the Belle's observation of an anomalously large $\Upsilon(nS)\pi^+
\pi^-,~n=1,2,3$ production around the $\Upsilon(5S)$ lead to the proposal of the
existence of a new resonance. A Breit-Wigner resonance shape fit yields a peak
mass of $(10888.4^{+2.7}_{-2.6}\pm1.2)~\MeV/c^2$, which is called 
$Y_b(10890)$ \cite{Chen:2008xia}. 

There are many theoretical interpretations for 
these  states \cite{Brambilla:2010cs,Nielsen:2009uh,Olsen:2009gi}.  
In the case of $Y(4260)$, although it seems not to fit the charmonium spectrum
\cite{pdg}, a proposal to accommodate it as a $4S$ state has been made 
in \cite{charm}. There are many other interpretations for this state: tetraquark 
state \cite{tetraquark}, hadronic molecule of $D_{1} D$, $D_{0} D^*$ 
\cite{Ding,Albuquerque}, $\chi_{c1} \omega$ \cite{Yuan}, $\chi_{c1} \rho$ 
\cite{liu}, $J/\psi f_0(980)$ \cite{oset}, a hybrid charmonium \cite{zhu}, a charm 
baryonium \cite{Qiao}, a cusp \cite{eef1,eef2,eef3}, etc. Within the available 
experimental information, none of these suggestions can be completely ruled out. 
For the $Y_b(10890)$ it has been interpreted as a tetraquark state, in a $P$-wave 
scalar-diquark scalar-antidiquark configuration \cite{ali,huang}. An alternative
scenario is that the anomalously large $\Upsilon(nS)\pi^+\pi^-,~n=1,2,3$ 
production observed by the Belle Collaboration does not come from a new resonance
but via sub-process $\Upsilon(5S)\to B^{(*)}B^{(*)}\to\Upsilon(1S,2S)\pi^+\pi^-$
\cite{meng,simo}.

In this work we use the QCD sum rule approach (QCDSR) \cite{svz,rry,SNB}
to check if the proposed $J/\psi f_0(980)$ assignment for the $Y(4260)$ \cite{oset}
is supported by a direct QCDSR calculation. We also study if a similar 
$\Upsilon f_0(980)$ current, and a tetraquark current (in a scalar-vector diquark 
configuration) could describe the $Y_b(10890)$ state.

\section{QCD sum Rules}
The QCDSR approach is based on the  two-point correlation function
\beq
\Pi_{\mu\nu}(q)=i\int d^4x ~e^{iq.x}\lag 0| T [j_\mu(x)j_\nu^\dagger(0)] |0\rag,
\lb{2po}
\enq
where the current $j_\mu(x)$ contains all the information about the hadron of
interest, like quantum numbers, quarks contents and so on.

We can write the correlation function in Eq.~(\ref{2po}) in terms of two 
independent Lorentz structures:
\beq
\Pi_{\mu\nu}(q)=-\Pi_1(q^2)(g_{\mu\nu}-{q_\mu q_\nu
\over q^2})+\Pi_0(q^2){q_\mu q_\nu\over q^2}.
\lb{lorentz}
\enq
The two invariant functions, $\Pi_1$ and $\Pi_0$, appearing in
Eq.~(\ref{lorentz}), have respectively the quantum numbers of the spin 1
and 0 mesons. Therefore, we choose to work with the Lorentz structure 
$g_{\mu\nu}$, since it gets contributions only from the vector state.

The QCD sum rule is obtained by evaluating the correlation function in 
Eq.~(\ref{2po}) in two ways: in the OPE side, we calculate the correlation 
function at the quark level in terms of quark and gluon fields.  We work at 
leading order in $\alpha_s$ in the operators, we consider the contributions 
from condensates up to dimension six and we keep terms which are linear 
in the strange quark mass $m_s$. In the phenomenological side,
the correlation function is calculated by inserting intermediate states 
for the hadronic state, H, and parametrizing the coupling of these states to the 
current $j_\mu(x)$, in terms of a generic coupling parameter $\lambda$, so that:
\beq
  \lag 0 | j_\mu| Y \rag =  \lambda \:\varepsilon_\mu,
\enq
where $\varepsilon_\mu$ is the polarization vector.

The phenomenological side of Eq.~(\ref{2po}), in the $g_{\mu\nu}$ structure, 
can be written as 
\beq
\Pi_1^{phen}(q^2)={\lambda^2\over M_{_H}^2-q^2} + \int\limits_{0}^\infty ds\, 
{\rho^{cont}(s)\over s-q^2}, \lb{phe} 
\enq
where $M_{_H}$ is the hadron mass and the second term in the RHS of 
Eq.(\ref{phe}) denotes higher resonance contributions.
The correlation function in the OPE side can be written as a dispersion relation:
\beq
\Pi_1^{ope}(q^2)=\int\limits_{4m_Q^2}^\infty ds {\rho^{ope}(s)\over s-q^2}\;,
\lb{ope}
\enq
where $m_Q$ is the heavy quark mass and $\rho^{ope}(s)$ is given by the 
imaginary part of the correlation function: $\pi \rho^{ope}(s)=\mbox{Im}
[\Pi_1^{ope}(s)]$. 

As usual in the QCD sum rules method, it is assumed that the continuum 
contribution to the spectral density, $\rho^{cont}(s)$ in Eq.~(\ref{phe}), 
vanishes 
bellow a certain continuum threshold $s_0$. Above this threshold, it is given by
the result obtained with the OPE. Therefore, one uses the ansatz \cite{io1}
\beq
\rho^{cont}(s)=\rho^{ope}(s)\Theta(s-s_0)\;.
\enq
In general, the continuum threshold $s_0$ is a parameter of the calculation
which is connected to the mass of the studied state, $H$, by the relation 
$s_0 \sim (M_{_H} + 0.5 \GeV)^2$.

To improve the matching between the two sides of the sum rule, we 
perfom a Borel transformation, which introduces the Borel parameter 
$\tau=1/M^2$, where $M$ is the Borel mass.
After transferring the continuum contribution to the OPE side the sum rule,
in the $g_{\mu\nu}$ structure, can be written as
\beq 
  \lambda^2e^{-M^2_{_H} \:\tau}=\int\limits_{4m_Q^2}^{s_0}ds~
  e^{-s \:\tau}~\rho^{ope}(s)\;.
  \label{sr1}
\enq
To extract $M_{_H}$ we take the derivative of Eq.~(\ref{sr1}) with respect to Borel
parameter $\tau$ and divide the result by Eq.~(\ref{sr1}), so that:
\beq 
  M^2_{_H} = \frac{\int\limits_{4m_Q^2}^{s_0}ds~ s \:e^{-s \:\tau}~\rho^{ope}(s)}
  {\int\limits_{4m_Q^2}^{s_0}ds~ e^{-s \:\tau}~\rho^{ope}(s)}\;.
  \label{mass}
\enq

\section{\boldmath $J/\psi \:f_0(980)$ Molecular State}

A possible current that couples with a $J/\psi \:f_0(980)$ molecular state, with
the quantum numbers $J^{PC}=1^{--}$, is given by:
\beq
  j_\mu = \left( \bar{c}_i \:\gamma_\mu \:c_i \right)\left( \bar{s}_j\:s_j\right)
  \label{singlet}
\enq
where $i,j$ are color indices and $c_i, s_j$ are the charm and strange quark 
fields respectively. Although there are conjectures that the $f_0(980)$
itself could be a tetraquark state \cite{jaffe}, in ref.~\cite{tetrarev} it
was shown that it is difficult to explain the light scalars as tetraquark states
from a QCDSR calculation. Therefore, here we use a simple quark-antiquark
current to describe the $f_0(980)$.

Another possibility for the current is considering  the vector and scalar parts 
in a color octet configuration: 
\beq
  j^{_{\lambda}}_\mu = \left( \bar{c}_i\:\lambda_{ij}^{_A}\gamma_\mu \:c_j \right)
  \left( \bar{s}_l \:\lambda_{lk}^{_A} \:s_k \right) \;,
\label{octet}
\enq
where  $\lambda^{_A}$ are the Gell-Mann matrices. 
The two currents can be related by the change: $j^{_{\lambda}}_\mu\to j_\mu$ 
 with  $\lambda_{ij}^{_A} \rightarrow \delta_{ij}$. Although the current in 
Eq.~(\ref{octet}) can not be interpreted as a meson-meson current, since
the vector and scalar parts carry colour, for simplicity we  still
call it a molecular current. 
Since the currents in Eqs.~(\ref{singlet}) and (\ref{octet}) have the
lowest dimension for a four-quark current with the $1^{--}$ quantum numbers,
from the theory of composite-operator renormalization \cite{dixon} we expect 
these currents to be multiplicatively renormalizable.

The spectral density $\rho^{ope}(s)$, for the $J^{PC} = 1^{--}$ exotic state 
described by a $J/\psi \:f_0(980)$ molecular current,  
up to dimension-six condensates, can be written as:

\beqa
  \rho^{ope}(s) &=& \rho^{pert}(s)+\rh^{\qq}(s)+\rh^{\lag G^2\rag}(s) \nn \\
  && + \rh^{\mix}(s)+\rh^{\qq^2}(s).  ~~~~~
  \lb{rhoeq}
\enqa
The expressions for $\rho^{ope}(s)$ for the currents in Eqs.~(\ref{singlet}) 
and (\ref{octet}), using factorization hypothesis, are given  in 
appendix \ref{App1}.

To extract reliable results form the sum rule is necessary to stablish the 
Borel window. A valid sum rule exist when one can find a Borel window where there 
is a OPE  convergence, a $\tau$-stability and where there is a dominance of the
ground state contribution. The maximum value of $\tau$ parameter is determined by 
imposing that the contribution of the higher dimension condensate is smaller 
 than 20\% of the total contribution: 
$\tau_{max}$ is such that
\beq
\left|{\mbox{OPE summed up dim n-1 }(\tau_{max})\over\mbox{total 
contribution }(\tau_{max})}\right|=0.8.
\label{mmin}
\enq

Since the continuum 
contribution decreases with $\tau$, due to the dominance of the perturbative 
contribution, the minimum value of $\tau$ is determined by 
imposing that the ground state contribution is equal to the continuum 
contribution. To guarantee a reliable result extracted from sum rules it is
important that there is a $\tau$ stability inside the Borel window. 

For a consistent comparison with the results obtained for the other molecular
states using the QCDSR approach, we have considered here the same values 
used for the quark masses and condensates as in 
refs.~\cite{x3872,molecule,lee,bracco,Albuquerque,z12,zwid,narpdg}, listed 
in  Table \ref{Param}.

{\small
\begin{table}[t]
\setlength{\tabcolsep}{1.25pc}
\caption{QCD input parameters.}
\begin{tabular}{ll}
&\\
\hline
Parameters&Values\\
\hline
$m_b(m_b)$ & $(4.24 \pm 0.05) \GeV$ \\
$m_c(m_c)$ & $(1.23 \pm 0.05) \GeV$ \\
$m_s$ & $(0.13 \pm 0.03)\GeV$ \\
$\qq$ & $-(0.23 \pm 0.01)^3\GeV^3$\\
$\lag g_s^2 G^2 \rag$ & $0.88~\GeV^4$\\
$\kappa \equiv \qqs / \qq$ & $(0.74 \pm 0.03)$\\
$m_0^2 \equiv \mixs / \qqs$ & $0.8\GeV^2$\\
\hline
\end{tabular}
\label{Param}
\end{table}}

We start with the current in Eq.~(\ref{singlet}). As mentioned above,
the continuum threshold is a physical parameter that should be determined from
the spectrum of the mesons. The value of the continuum threshold in the QCDSR 
approach is, in general, given as the value of the mass of the first excited 
state squared. In some known cases, like the $\rho$ and $J/\psi$, 
the first excitated state has a mass approximately $0.5~\GeV$ above the
ground state mass.  In the cases that one does not know the spectrum, one 
expects the continuum threshold to be approximately the square of the mass 
of the state plus $0.5~\GeV$: $s_0=(M_H+0.5~\GeV)^2$. Therefore, to fix the 
continuum threshold range we extract the mass from the sum rule, for a given 
$s_0$, and accept such value of $s_0$ if the obtained mass is in the range 
0.4 GeV to 0.6 GeV smaller than $\sqrt{s_0}$. Using this criterion,
we obtain $s_0$ in the range $5.0\leq\sqrt{s_0}\leq 5.2~\GeV$.
\begin{figure}[t]
\begin{center}
{\begin{flushleft} a) \end{flushleft}} \vspace{-0.3cm}
\includegraphics[width=6.0cm]{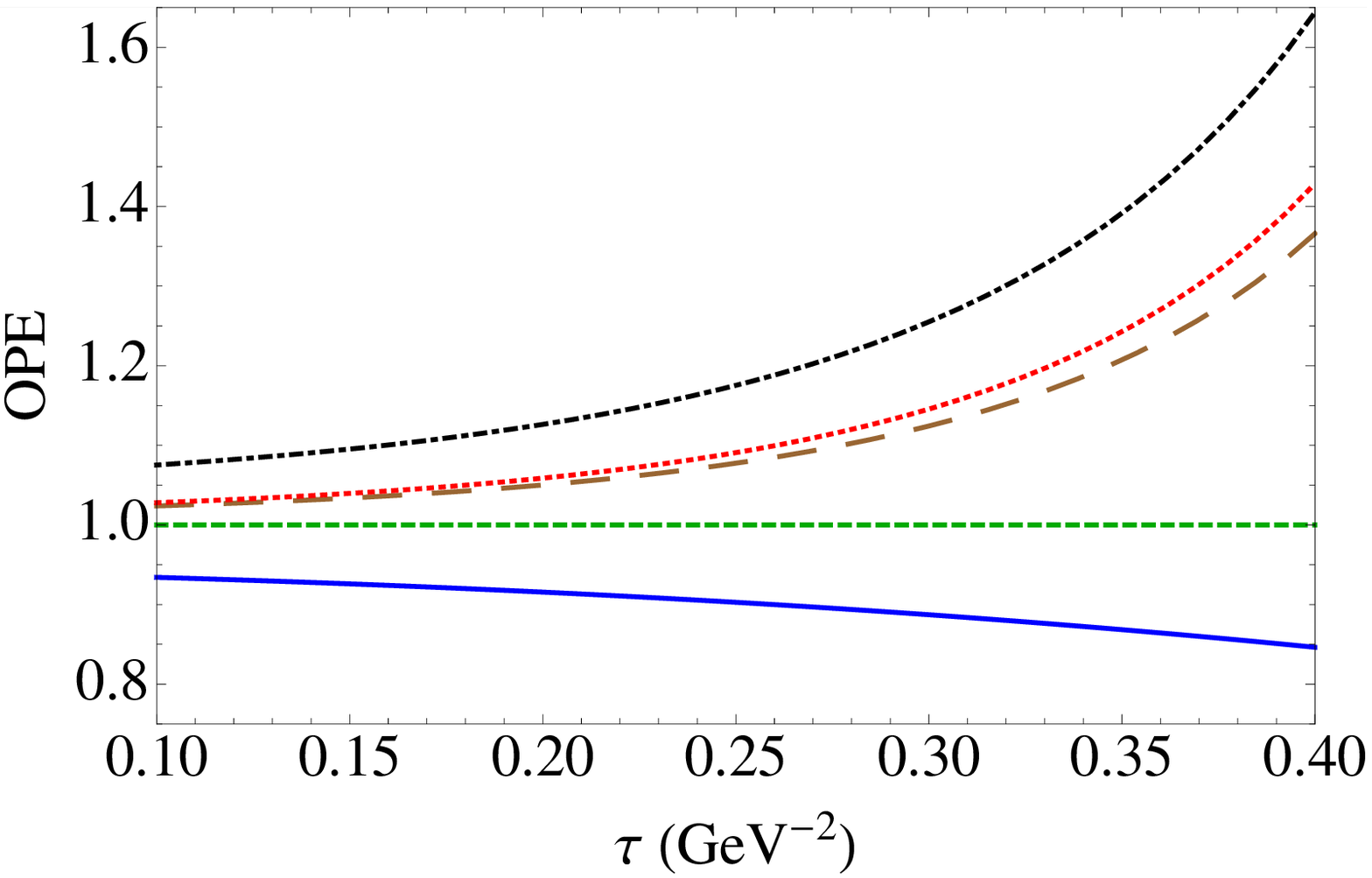}
{\begin{flushleft} b) \end{flushleft}} \vspace{-0.3cm}
\includegraphics[width=6.0cm]{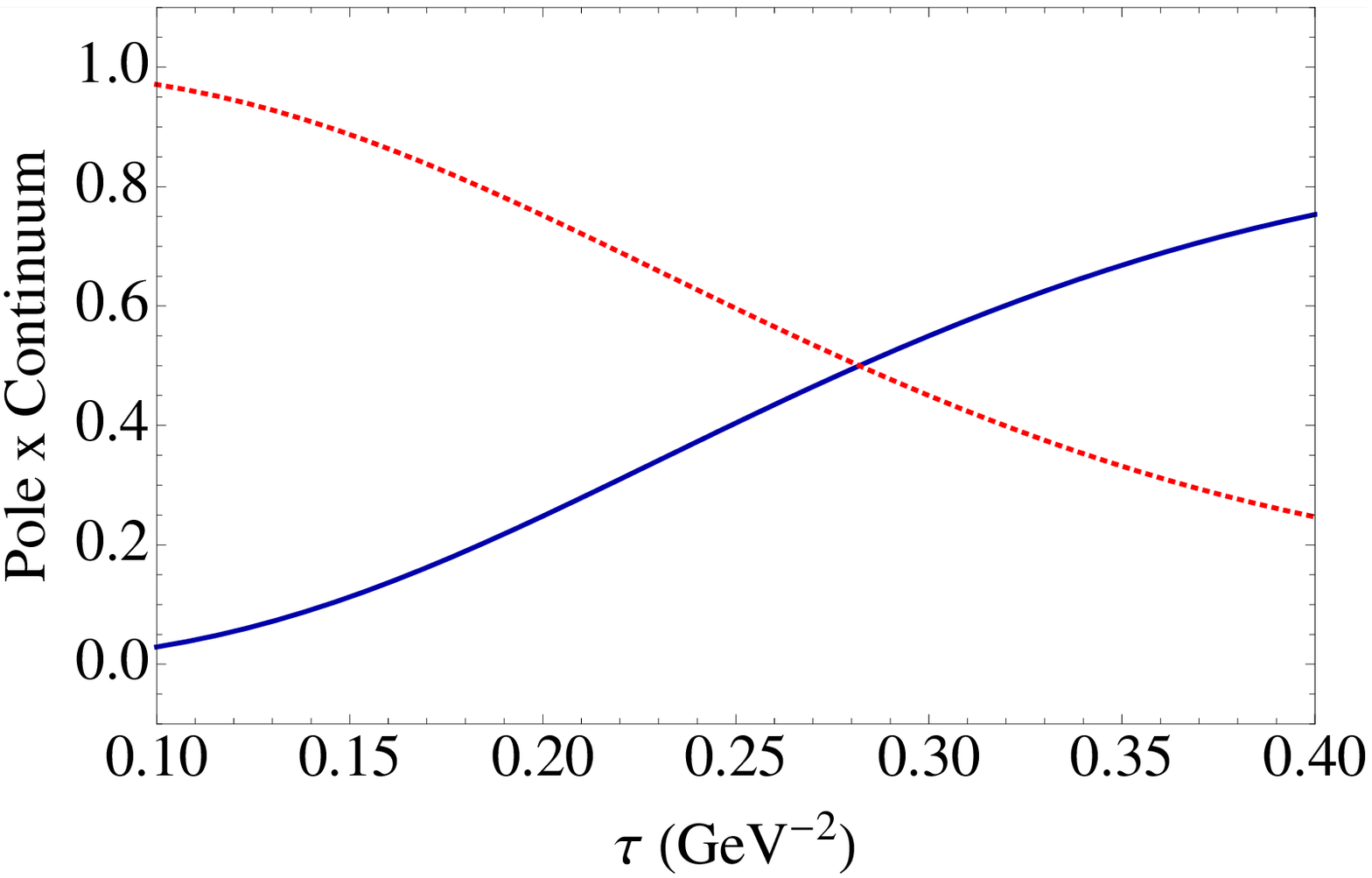}
{\begin{flushleft} c) \end{flushleft}} \vspace{-0.3cm}
\includegraphics[width=6.0cm]{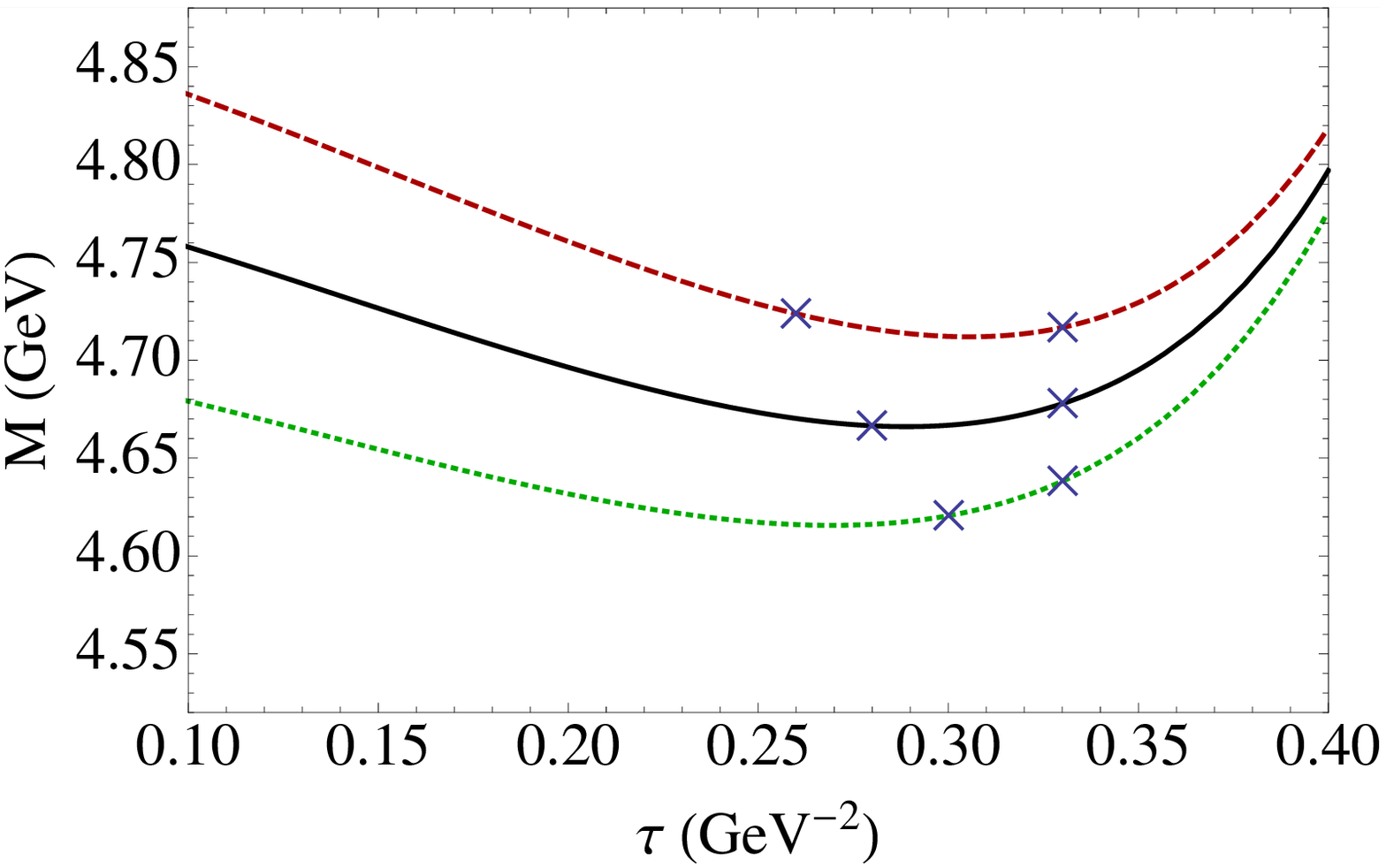}
\caption{\footnotesize $J/\psi \:f_0(980)$ current in a color singlet 
configuration. 
{\bf a)} OPE convergence in the region $0.10 \leq \tau \leq 0.40~\GeV^{-2}$ for 
$\sqrt{s_0} = 5.10 \GeV$.  We plot the 
relative contributions starting with the perturbative contribution 
(dot-dashed line), and each other line represents the 
relative contribution after adding of one extra condensate in the expansion: 
+ $\qqs$ (solid line), + $\langle G^2 \rangle$ (long-dashed line), + 
$\mixs$ (dotted line) and + $\qqs^2$ (dashed line).
{\bf b)} The pole (solid line) and continuum (dotted line) contributions for
$\sqrt{s_0} = 5.10 \GeV$. 
{\bf c)} The mass as a function of the sum rule parameter $\tau$ for 
$\sqrt{s_0} = 5.00 \GeV$ (dotted line), $\sqrt{s_0} = 5.10 \GeV$ (solid line) and
$\sqrt{s_0} = 5.20 \GeV$ (dot-dashed line). The crosses indicate the valid Borel 
window.}
\label{FigSinglet}
\end{center}
\end{figure}

In Fig.~\ref{FigSinglet}a) we show the relative contribution of the terms in the
OPE side of the sum rule, for $\sqrt{s_0} = 5.10~\GeV$. From this figure we see  
that the contribution of the dimension-6 condensate is smaller than 20\% of the 
total contribution for values of $\tau \leq 0.33 \GeV^{-2}$, which indicates a good 
OPE convergence. 
From Fig.~\ref{FigSinglet}b), we also see that the pole contribution is bigger 
than the continuum contribution only for 
$\tau \geq 0.28 \GeV^{-2}$. Therefore, we fix the Borel window as: 
$(0.28 \leq \tau \leq 0.33) \GeV^{-2}$. 
From  Eq.~(\ref{mass}), we can calculate the ground state mass, which is show,
as a function of $\tau$ , in the Fig.~\ref{FigSinglet}c). From this figure we see
that there is a very good $\tau$ stability in the determined Borel window, which is
shown, through the crosses, in Fig.~\ref{FigSinglet}c).

Varying the value of the continuum threshold in the range 
$\sqrt{s_0} = 5.10 \pm 0.10 \GeV$, and the other parameters as indicated in
Table \ref{Param} we get:
\beq
 M_Y = (4.67\pm0.09)~\GeV ~.
 \label{Msinglet}
\enq
This mass is not compatible with the  proposition in \cite{oset}, 
which describes the $Y(4260)$ state as the $J/\psi \:f_0(980)$ molecular state.
By the other hand, this result is in an excellent agreement with the mass of the 
$Y(4660)$ state. The obtained mass is largely above the $J/\psi~f_0(980)$ threshold
and, therefore, such molecular state would not be bound. One has to remember, 
however, that the current in Eq~(\ref{singlet}) is written in terms of the currents
that couples with the $J/\psi$ and $f_0(980)$ mesons, but it also couples with
all excited states with the  $J/\psi$ and $f_0(980)$ quantum numbers. From the
QCDSR analysis presented here we can only warranty that the mass in 
Eq.~(\ref{Msinglet}) is the mass of the ground state of all states 
described by the current in Eq~(\ref{singlet}), but not that its constituents, 
described by the $\bar{c}_i\gamma_\mu c_i$ and $\bar{s}_js_j$ currents, are the 
ground states of these currents: the $J/\psi$ and $f_0(980)$ mesons. Therefore, 
it is possible that the mass obtained in Eq~(\ref{Msinglet}) describes a 
$\psi'~f_0(980)$ molecular state,  since the $\psi'~f_0(980)$ threshold is at 
$4.66~\GeV$, compatible with a loosely bound state. The interpretation
of the $Y(4660)$ as a  $\psi'~f_0(980)$ molecular state was first proposed
in ref.~\cite{guo} and is also in agreement with the $Y(4660)$
main decay channel: $Y(4660)\to \Psi(2S)~\pi^+\pi^-$. It is also important to 
mention that our result indicates that, from a QCDSR point of view, there is
no $J/\psi~f_0(980)$ bound state.

\begin{figure}[t]
\begin{center}
{\begin{flushleft} a) \end{flushleft}} \vspace{-0.3cm}
\includegraphics[width=6.0cm]{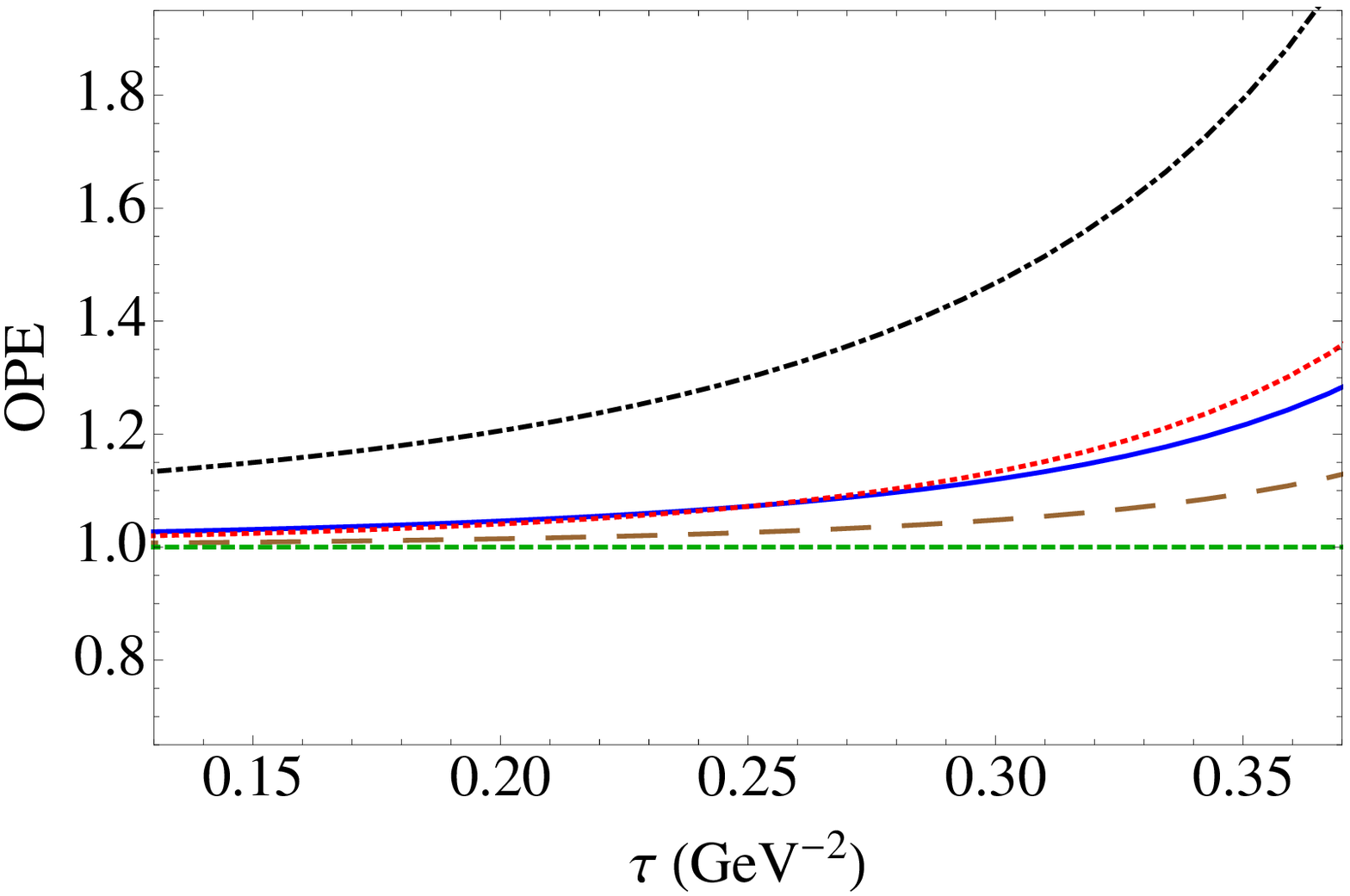}
{\begin{flushleft} b) \end{flushleft}} \vspace{-0.3cm}
\includegraphics[width=6.0cm]{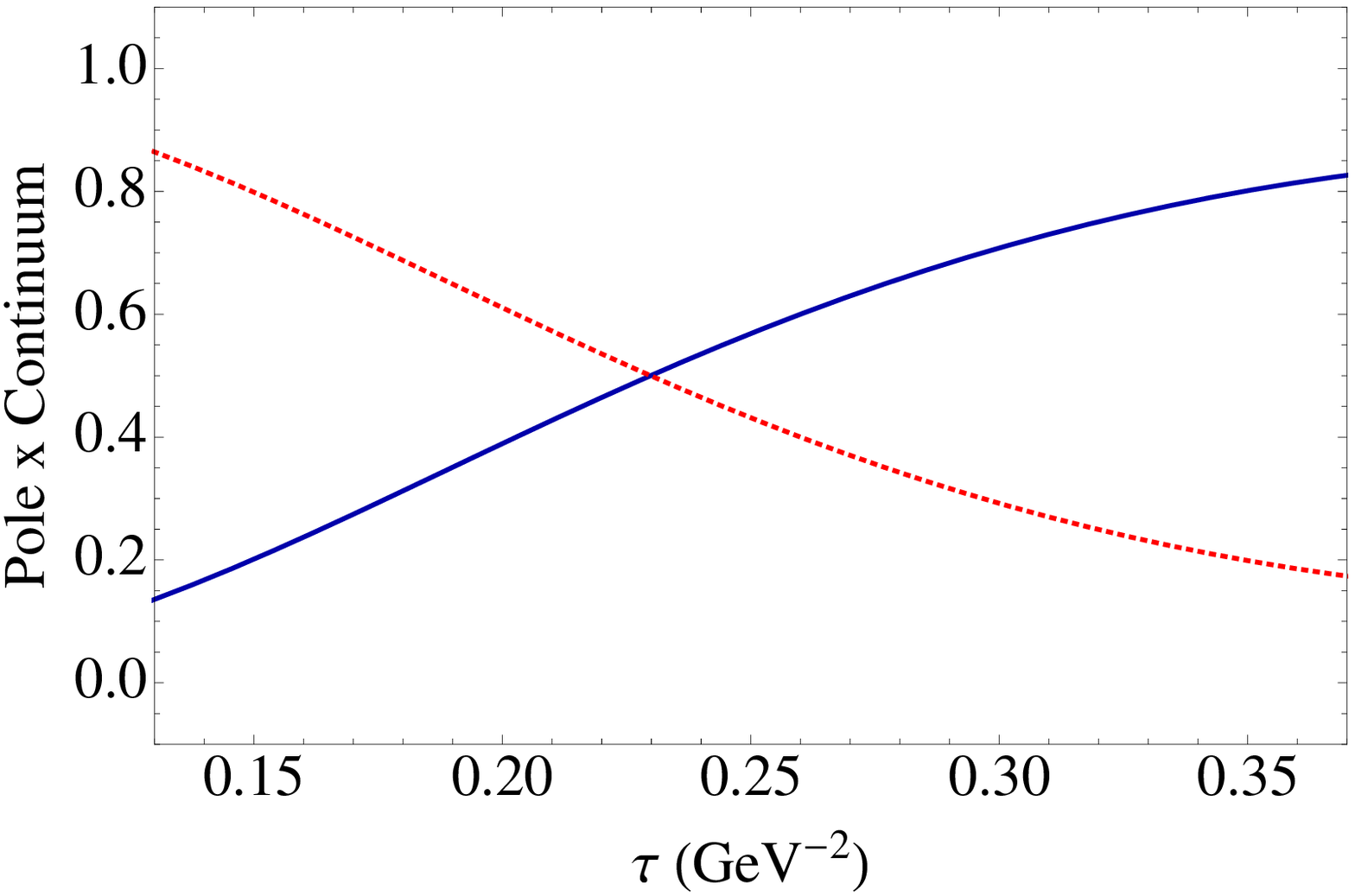}
{\begin{flushleft} c) \end{flushleft}} \vspace{-0.3cm}
\includegraphics[width=6.0cm]{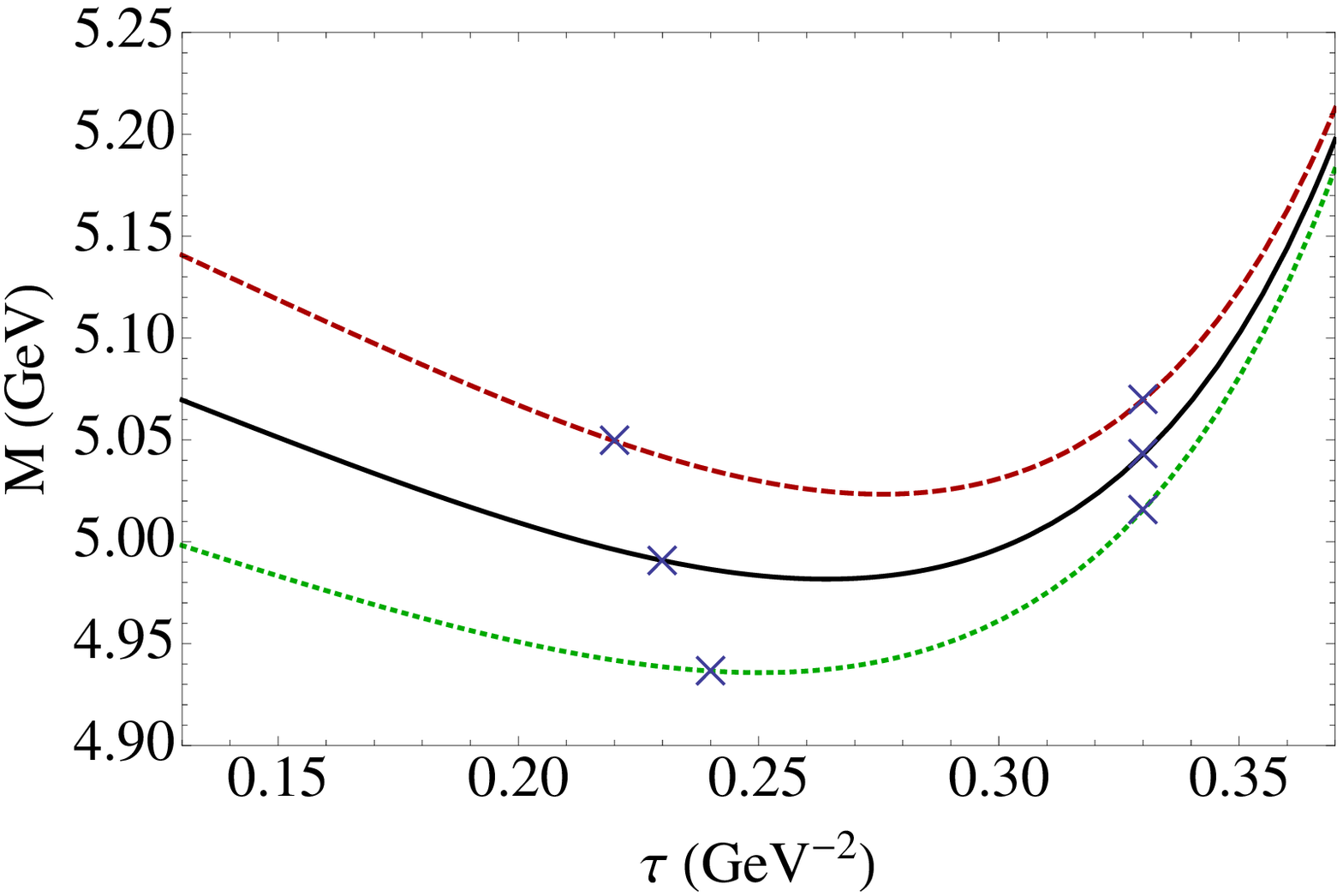}
\caption{\footnotesize $J/\psi \:f_0(980)$ molecule in a color octet configuration. 
{\bf a)} OPE convergence in the region $0.13 \leq \tau \leq 0.37~\GeV^{-2}$ for 
$\sqrt{s_0} = 5.50 \GeV$.  We plot the 
relative contributions starting with the perturbative contribution 
(dot-dashed line), and each other line represents the 
relative contribution after adding of one extra condensate in the expansion: 
+ $\qqs$ (solid line), + $\langle G^2 \rangle$ (long-dashed line), + 
$\mixs$ (dotted line) and + $\qqs^2$ (dashed line).
{\bf b)} The pole (solid line) and continuum (dotted line) contributions for
$\sqrt{s_0} = 5.50 \GeV$. 
{\bf c)} The mass as a function of the sum rule parameter $\tau$ for 
$\sqrt{s_0} = 5.40 \GeV$ (dotted line), $\sqrt{s_0} = 5.50 \GeV$ (solid line) and
$\sqrt{s_0} = 5.60 \GeV$ (dot-dashed line). The crosses indicate the valid Borel 
window.}
\label{FigOctet}
\end{center}
\end{figure}

In the case of the current in Eq.~(\ref{octet})
we obtain $s_0$ in the range $5.4\leq\sqrt{s_0}\leq 5.6~\GeV$. The results 
for this current are shown in Fig.~\ref{FigOctet}, from where we can see that,
for $\sqrt{s_0} = 5.50 \GeV$, the Borel 
window is fixed as: $(0.23 \leq \tau \leq 0.33) \GeV^{-2}$. Varying the continuum 
threshold in the range $\sqrt{s_0} = 5.50 \pm 0.10 \GeV$, and the other parameters 
as indicated in Table \ref{Param} we get:
\beq
 M_{Y_{\lambda}} = (5.00\pm0.10)~\GeV ~.
 \label{Moctet}
\enq
This value for the mass is not compatible with any observed charmonia state.
Besides, comparing the results in Eqs.~(\ref{Msinglet}) and (\ref{Moctet}), 
we conclude that a molecular state with $\bar{c}\gamma_\mu c$ and $\bar{s}s$ in 
color octet configurations has a bigger mass than the similar state when 
$\bar{c}\gamma_\mu c$ and $\bar{s}s$
are in a color singlet configuration. This result is the opposite than the
result obtained in ref.~\cite{xcurrents} for a $J/\psi\pi$ current. However,
in ref.~\cite{xcurrents} the same range of the continuum threshold was used
for both currents. In the present case we see that if we use $\sqrt{s_0} = 5.10 
\pm 0.10 \GeV$ for the current in Eq.~(\ref{octet}) we do not find a Borel window.
This is the reason why we had to work with bigger values of $s_0$ for the
current in Eq.~(\ref{octet}).

\section{\boldmath $J/\psi \sigma(600)$ Molecular State}

It is straightforward to extend the study presented in the above section for
the non-strange case. To do that one only has to use $\sss=\qq$ and $m_s=0$ in the
spectral density expressions given in appendix \ref{App1}. In this case, to obtain 
a Borel window we need to release the condition in Eq.~(\ref{mmin}) and allow
$\tau_{max}$ to be determine by imposing that the dimension-6 condensate
could be 25\% of the total contribution. This indicates that the OPE convergence
is worse in this case as compared with the $J/\psi f_0(980)$ case. This is
due to the fact that the dimension-3 and dimension-5 condensates do not contribute
in this case. In Table \ref{psi-sigma} we present the result 
obtained for the mass with the two currents, together with the used
continuum threshold range.

{\small
\begin{table}[h,t,b]
\setlength{\tabcolsep}{1.0pc}
\caption{Results for the $J/\psi~\sigma$ currents.}
\begin{tabular}{ccc}
&\\
\hline
current in Eq.&$M_H(\GeV)$&$\sqrt{s_0}(\GeV)$\\
\hline
(\ref{singlet}) & $4.63 \pm 0.10$& $5.1\pm0.1$ \\
(\ref{octet}) & $4.97 \pm 0.08$& $5.5\pm0.1$ \\
\hline
\end{tabular}
\label{psi-sigma}
\end{table}}

As one can see from Table \ref{psi-sigma}, the results obtained with the
$J/\psi ~\sigma(600)$ molecular current are in agreement with the results
obtained with the $J/\psi~ f_0(980)$ current, which is not what one would naively
expect. However, this kind of findings is not uncommon in QCDSR calculations
for multiquark states \cite{x3872}. Again, since the masses obtained are largely
above the $J/\psi ~\sigma(600)$ threshold, we conclude that there is no
$J/\psi ~\sigma(600)$ bound state. In this case, since the mass obtained is also
above the $\psi' ~\sigma(600)$ threshold we can not interpret the $Y(4660)$
as a  $\psi' ~\sigma(600)$ molecular state, despite the fact that the obtained mass
is in agreement with the $Y(4660)$ mass.

\section{\boldmath $\Upsilon f_0(980)$ and $\Upsilon \sigma(600)$ Molecular 
States}

It is also straightforward to extend the previous study to the b-sector. 
To do that one only has to make the change $m_c \rightarrow m_b$ 
and for the non-strange case use $m_s=0$ in the spectral density expressions 
given in the appendix \ref{App1}. This allow us to study the molecular currents: 
$\Upsilon \:f_0(980)$ and $\Upsilon \:\sigma(600)$, in both,
color singlet and color octet configuration, as given in Eqs.~(\ref{singlet})
and (\ref{octet}). We obtain similar OPE convergence and pole dominance as
in the charm sector. 

{\footnotesize
\begin{table}[h,t,b]
\setlength{\tabcolsep}{.35pc}
\caption{Results for the $\Upsilon~f_0$ and $\Upsilon~\sigma$ currents.}
\begin{tabular}{lccc}
&\\
\hline
States& $M_H$& Borel Window & $\sqrt{s_0}$\\
&($\GeV$)&($\GeV^{-2}$)&($\GeV$)\\
\hline
{\tiny Color Singlet}\\
$\Upsilon \:f_0(980)$ & $10.75 \pm 0.12$& $0.11 \leq \tau \leq 0.15$ &$11.3\pm0.1$ \\
$\Upsilon \:\sigma(600)$ & $10.74 \pm 0.09$& $0.11 \leq \tau \leq 0.13$& $11.3
\pm0.1$ \\
&&& \\
{\tiny Color Octet}\\
$\Upsilon \:f_0(980)$ &$11.08 \pm 0.11$& $0.11 \leq \tau \leq 0.14$& $11.7\pm0.1$ \\
$\Upsilon \:\sigma(600)$ & $11.09 \pm 0.10$& $0.10 \leq \tau \leq 0.13$& $11.7
\pm0.1$ \\
\hline
\end{tabular}
\label{upsilon}
\end{table}}

In Table \ref{upsilon}  we present the 
results obtained for the masses of the states described by $\Upsilon \:f_0(980)$ 
and  $\Upsilon \:\sigma(600)$ currents, together with their 
respective continuum threshold range and valid Borel window.

From this Table we see that, as in the charm sector, the masses obtained with 
the $\Upsilon~f_0$ and $\Upsilon~\sigma$ currents are very similar and that
the relative differences with the singlet and octet currents are smaller than
in the charm sector. This is also consistent with the findings
in ref.~\cite{zhu2}, where different $1^{--}$ tetraquark currents where used,
in a QCDSR calculation, with similar results for the different currents and also
for non-strange and strange sectors.
Considering the errors, all the masses obtained with these 
currents are compatible with the mass of the recently oserved $Y_b(10890)$ state. 
However, since the  $\Upsilon(1S)~f_0$  and $\Upsilon(2S)~f_0$ thresholds
are at 10.44 GeV and 11.00 GeV respectively, and that the thresholds with 
$\sigma$ are around 380 MeV below these numbers, the only possible molecular 
interpretation for the $Y_b(10890)$ is that it could be a $\Upsilon(2S)~f_0$ 
molecular state.

\section{\boldmath Tetraquark current for the $Y_b(10890)$}
%
\begin{figure}[t]
\begin{center}
{\begin{flushleft} a) \end{flushleft}} \vspace{-0.3cm}
\includegraphics[width=6.0cm]{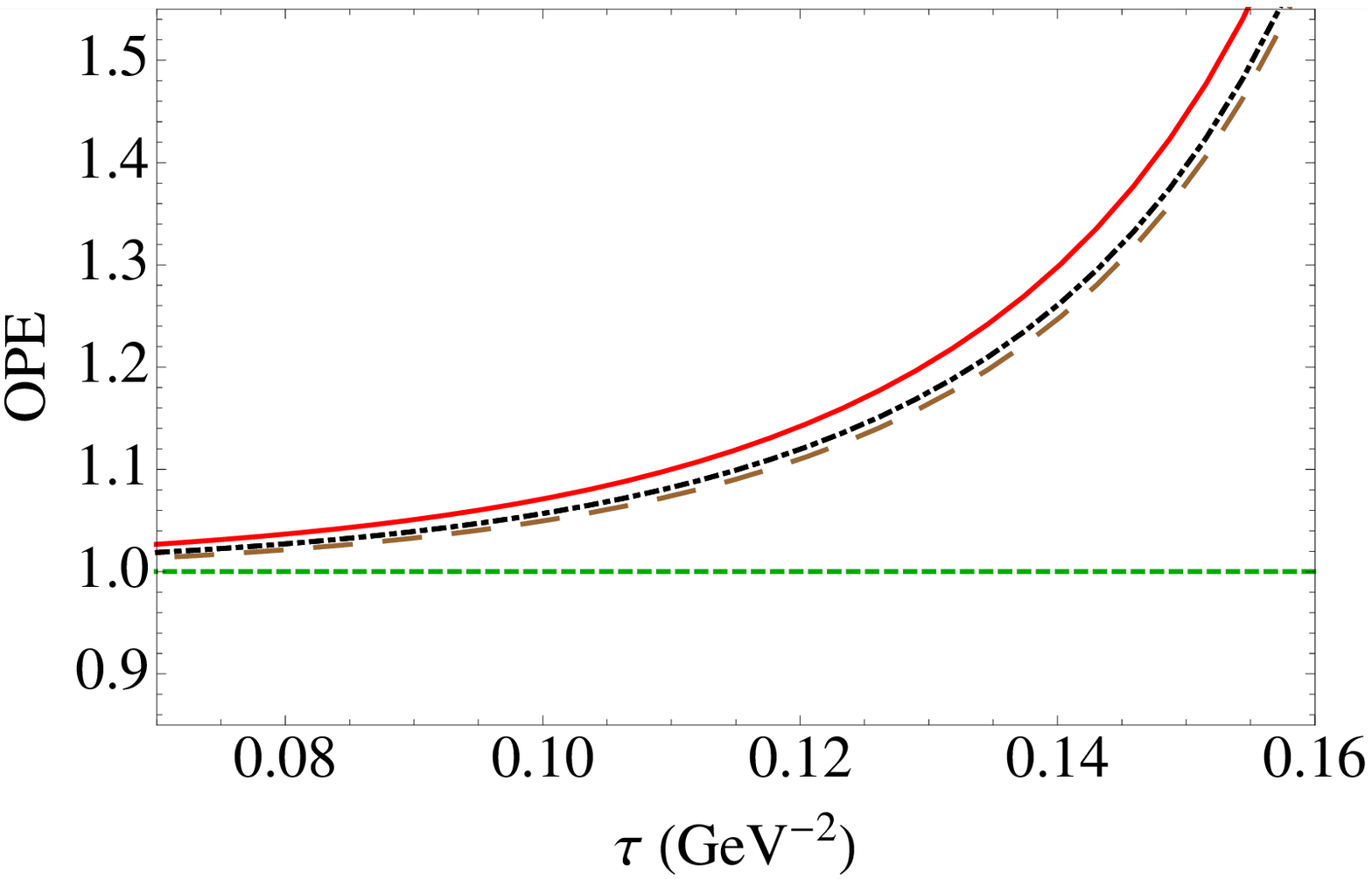}
{\begin{flushleft} b) \end{flushleft}} \vspace{-0.3cm}
\includegraphics[width=6.0cm]{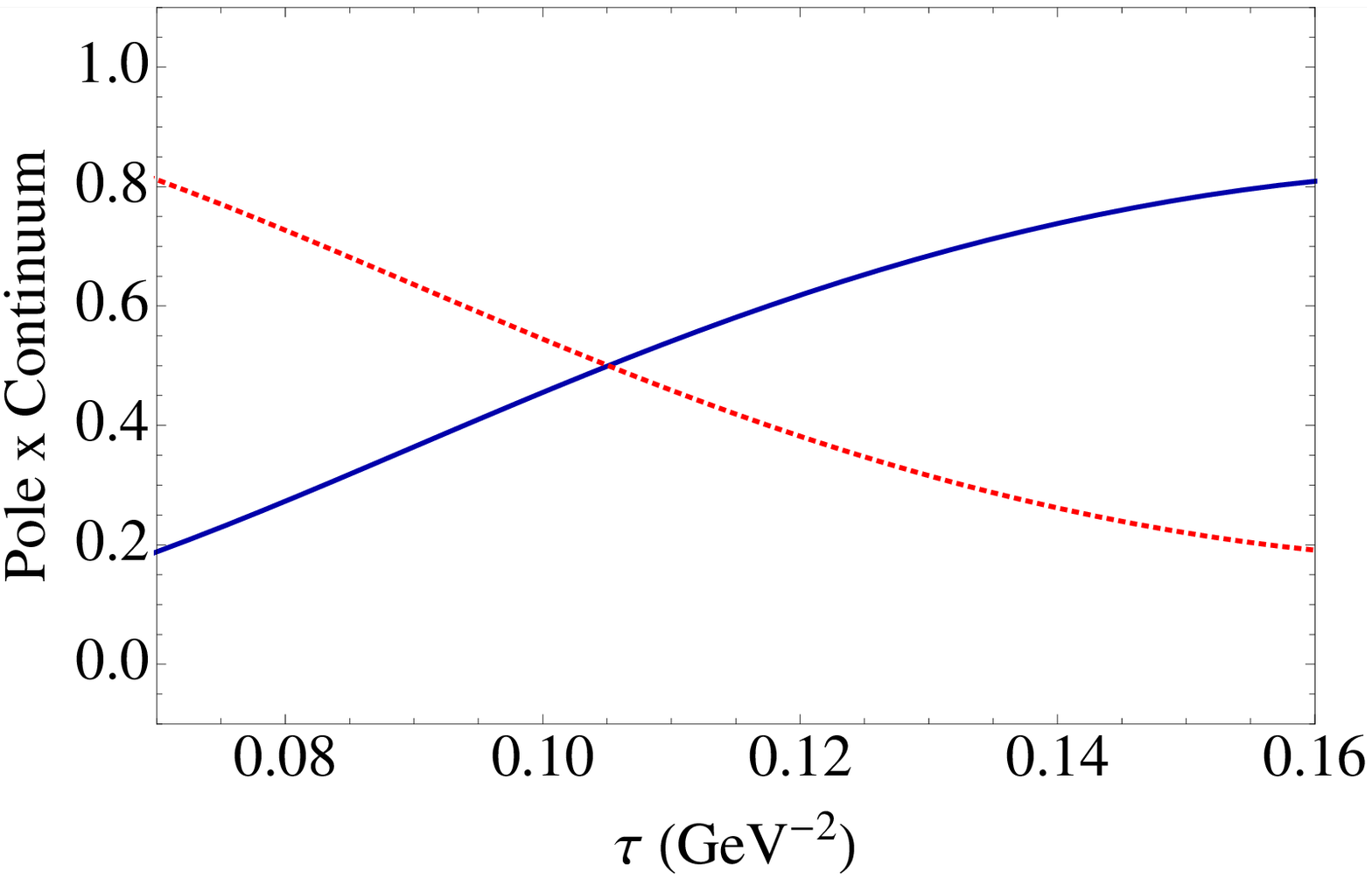}
{\begin{flushleft} c) \end{flushleft}} \vspace{-0.3cm}
\includegraphics[width=6.0cm]{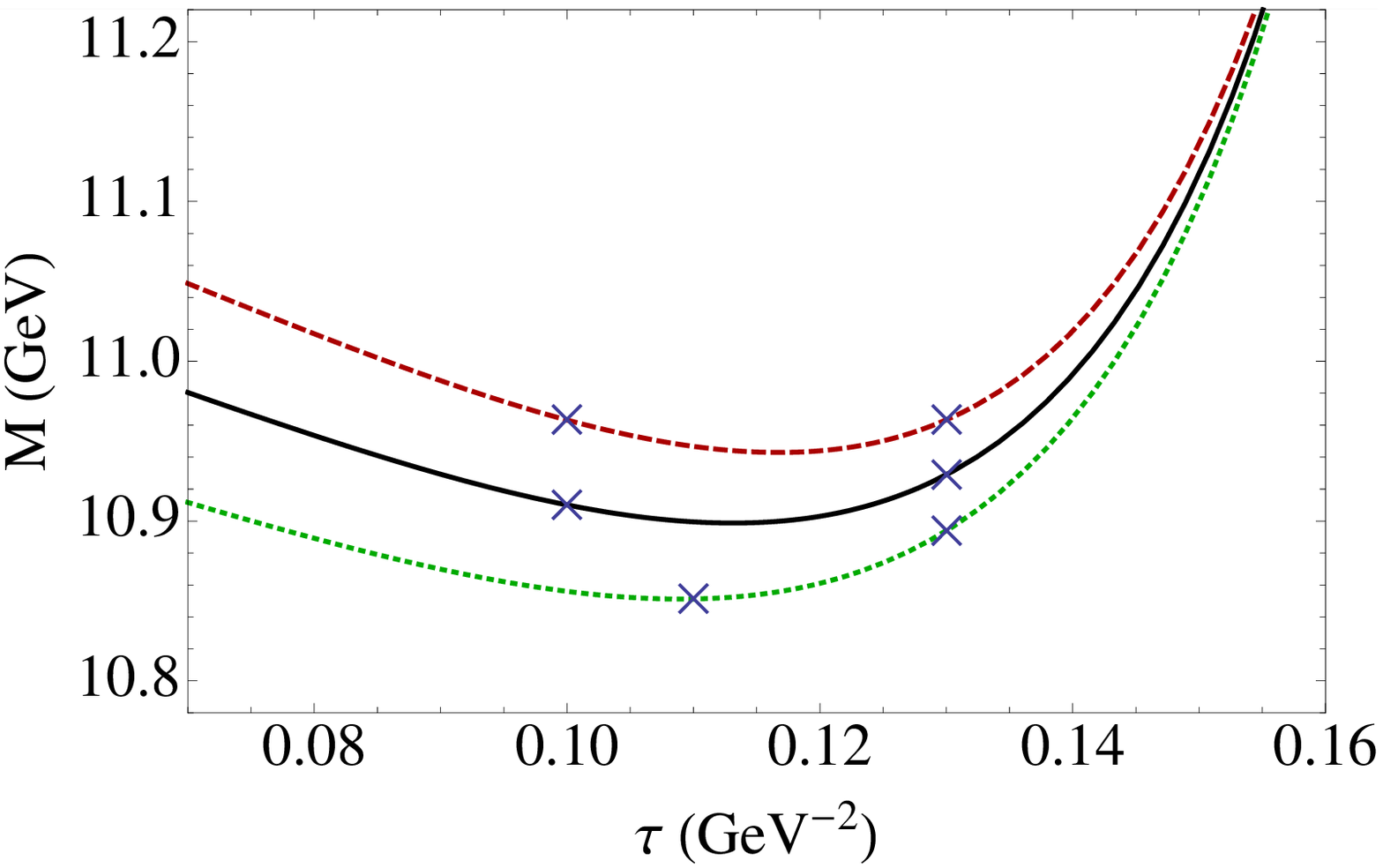}
\caption{\footnotesize $[bq][\bar{b}\bar{q}]$ tetraquark state.
{\bf a)} OPE convergence in the region $0.07 \leq \tau \leq 0.16~\GeV^{-2}$ for 
$\sqrt{s_0} = 11.50 \GeV$.  We plot the relative contributions starting with the 
perturbative 
contribution (dot-dashed line), and each other line represents the relative 
contributions after 
adding of one extra condensate in the expansion:
+ $\langle G^2 \rangle$ (long-dashed line), + $\mix$ (solid line)
and + $\qq^2$ (dashed line).
{\bf b)} The pole (solid line) and continuum (dotted line) contributions for
$\sqrt{s_0} = 11.50 \GeV$. 
{\bf c)} The mass as a function of the sum rule parameter $\tau$ for 
$\sqrt{s_0} = 11.40 \GeV$ (dotted line), $\sqrt{s_0} = 11.50 \GeV$ (solid line) and
$\sqrt{s_0} = 11.60 \GeV$ (dot-dashed line). The crosses indicate the valid Borel 
window.}
\label{FigYb}
\end{center}
\end{figure}

In ref.~\cite{ali} the $Y(10890)$ was interpreted as a bound tetraquark state
$[bq][\bar{b}\bar{q}]={\cal Q}\bar{\cal Q}$ with the spin and angular momentum
quantum numbers:  $S_{\cal Q}=0,~S_{\bar{\cal Q}}=0,~S_{{\cal Q}\bar{\cal Q}}=0,
~L_{{\cal Q}\bar{\cal Q}}=1$. This same configuration was used in a QCDSR
calculation \cite{huang} and the obtained mass was $10.88\pm0.13~\GeV$,
in a very good agreement with the $Y(10890)$ mass. In this section we want to 
check if the tetraquark current constructed with scalar and vector diquarks: 
\beqa
  j^{_{Y_{b}}}_\mu &=& {\epsilon_{ijk}\epsilon_{lmk}\over\sqrt{2}} 
\Bigg[(q_i^TC\gamma_5b_j)
  (\bar{q}_l\gamma_\mu\gamma_5 C\bar{c}_m^T) \nn \\
  && +(q_i^TC\gamma_5\gamma_\mu b_j) (\bar{q}_l\gamma_5C\bar{c}_m^T) \Bigg]
  \label{yb}
\enqa
can also be used to describe the $Y(10890)$. In Eq.~(\ref{yb}) $i,j,k,\ldots$ are 
color indices, $C$ is the charge conjugation matrix, $q = u,d,s$ is the 
light quark field and $b$ is the quark bottom field. Notice that the main decay 
channel: 
$Y_b \rightarrow \Upsilon(1S) \pi^+ \pi^-$, does not necessary indicate that $Y_b$
has only light-quarks in its composition, if it is interpreted as a four-quark state.
In fact, it is very interesting to investigate any possiblity to the quark content 
for this tetraquark state. The expressions for $\rho^{ope}(s)$ for the current in 
Eq.~(\ref{yb}) are given  in appendix \ref{App1}.

\begin{figure}[t]
\begin{center}
{\begin{flushleft} a) \end{flushleft}} \vspace{-0.3cm}
\includegraphics[width=6.0cm]{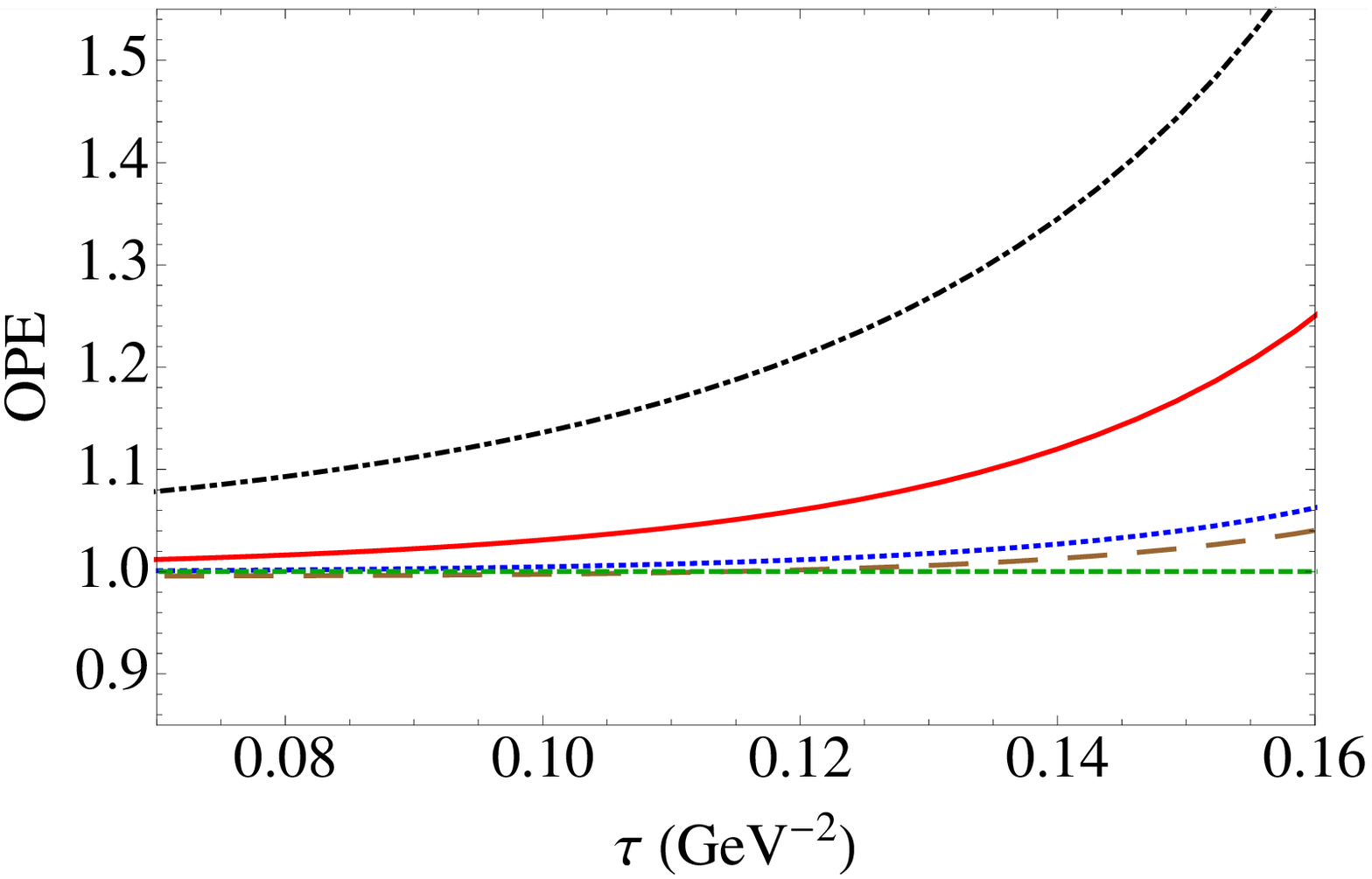}
{\begin{flushleft} b) \end{flushleft}} \vspace{-0.3cm}
\includegraphics[width=6.0cm]{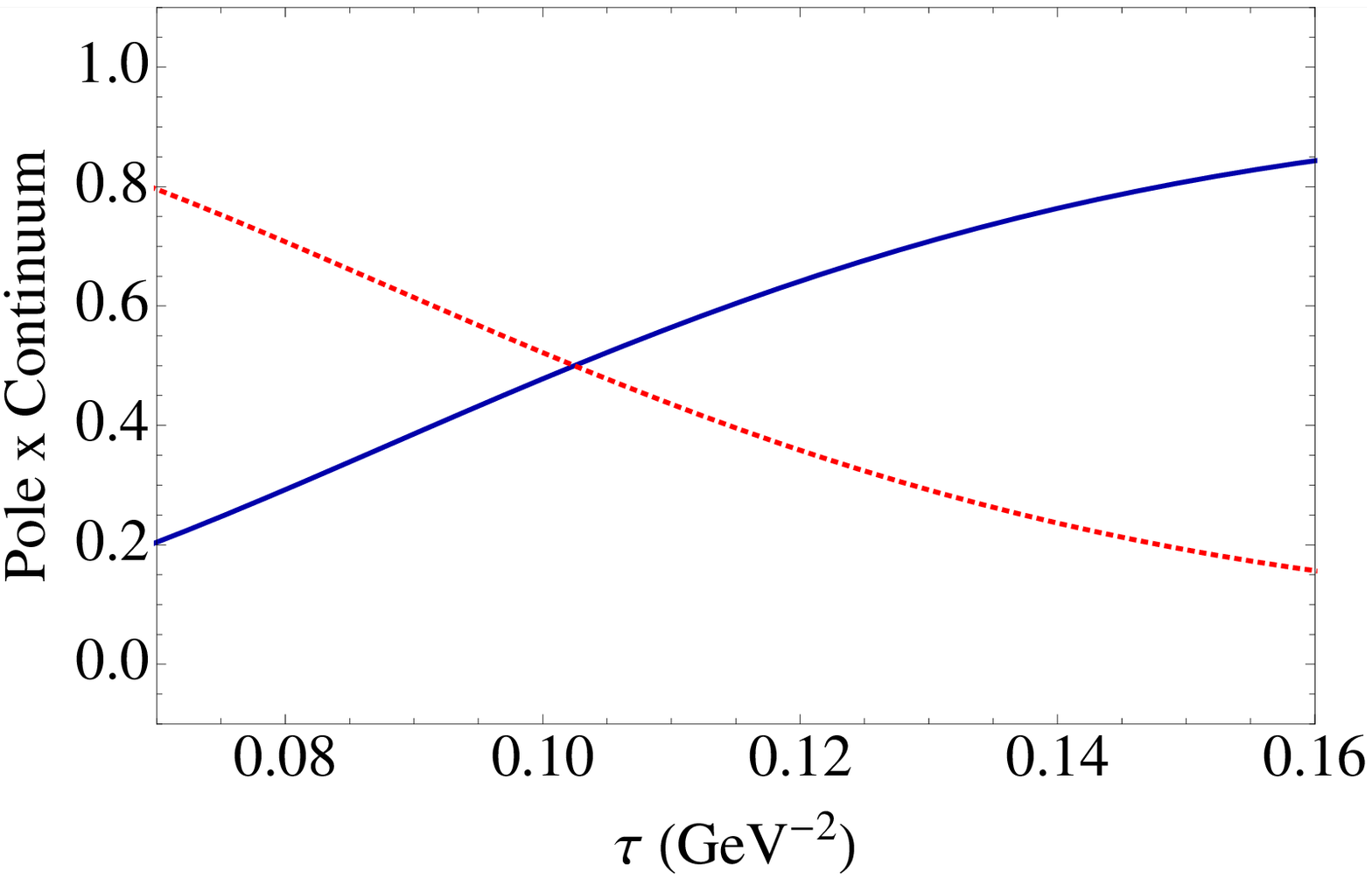}
{\begin{flushleft} c) \end{flushleft}} \vspace{-0.3cm}
\includegraphics[width=6.0cm]{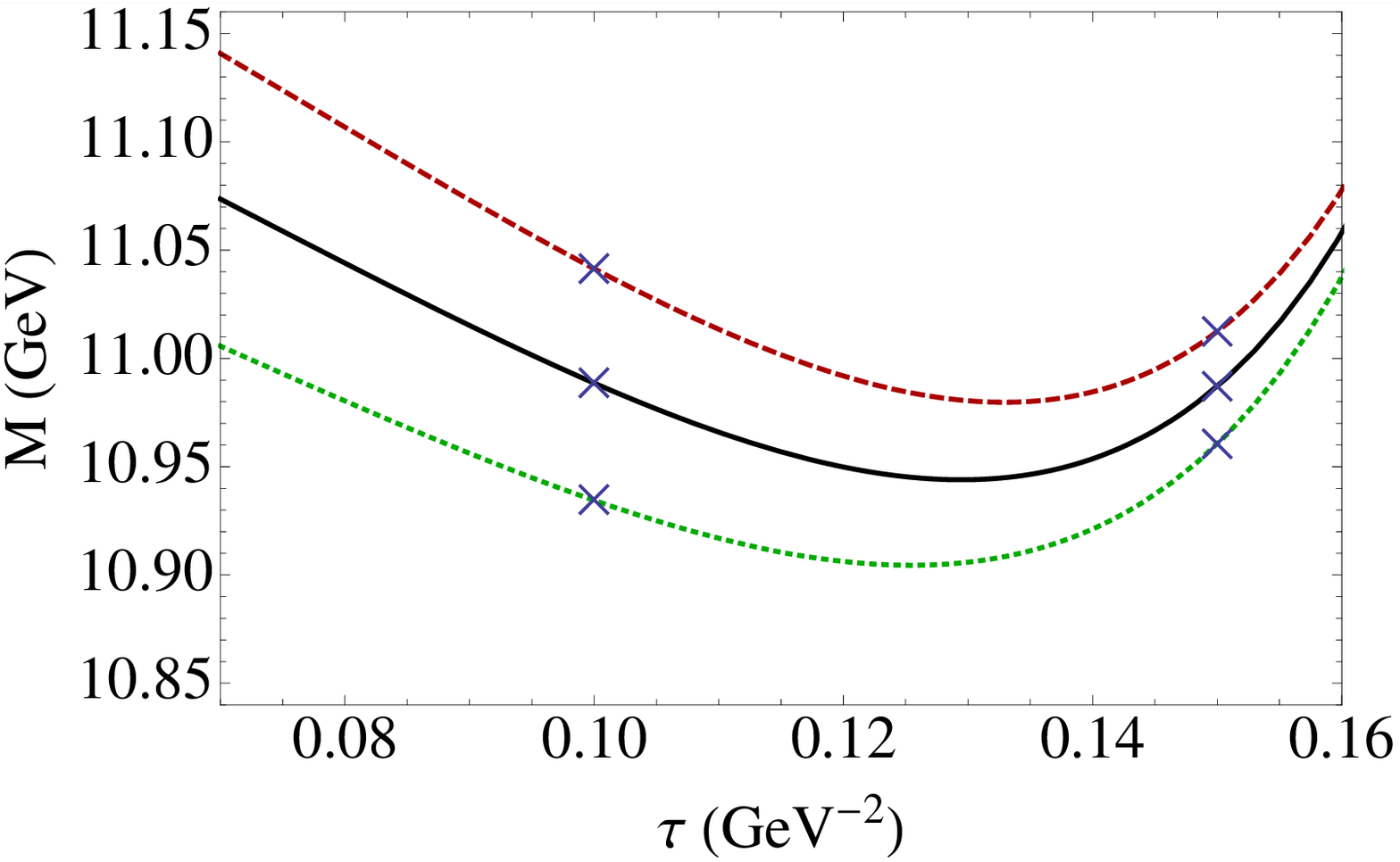}
\caption{\footnotesize $[bs][\bar{b}\bar{s}]$ tetraquark state.
{\bf a)} OPE convergence in the region $0.07 \leq \tau \leq 0.16~\GeV^{-2}$ for 
$\sqrt{s_0} = 11.60 \GeV$.  We plot the relative contributions starting with the 
perturbative 
contribution (dot-dashed line), and each term other line represents the relative 
contributions 
after adding of one extra condensate in the expansion:
+ $\qqs$ (dotted line), + $\langle G^2 \rangle$ (long-dashed line), $\mixs$ (solid 
line)
and $\qqs^2$ (dashed line).
{\bf b)} The pole (solid line) and continuum (dotted line) contributions for
$\sqrt{s_0} = 11.60 \GeV$. 
{\bf c)} The mass as a function of the sum rule parameter $\tau$ for 
$\sqrt{s_0} = 11.50 \GeV$ (dotted line), $\sqrt{s_0} = 11.60 \GeV$ (solid line) and
$\sqrt{s_0} = 11.70 \GeV$ (dot-dashed line). The crosses indicate the valid Borel 
window.}
\label{FigYbs}
\end{center}
\end{figure}

We consider first the $[bq][{\bar{b}\bar{q}}]$ tetraquark state and the results are 
shown in Fig.~\ref{FigYb}. As explained above, we extract the mass from the sum 
rule, 
for a given $s_0$, and accept such value if the obtained mass is around $\sqrt{s_0}-
0.5~\GeV$. Using this criteria we got $11.4~\GeV\leq\sqrt{s_0}\leq11.6~\GeV$.

Varying the value of the continuum threshold in the range: 
$\sqrt{s_0} = 11.50 \pm 0.10 \GeV$, and taking into account the uncertainties 
as indicated in Table \ref{Param} we get:
\beq
 M_{Y_{b}} = (10.91\pm0.07)~\GeV ~,
\label{ybq}
\enq
which is in an excellent agreement with the observed 
mass for $Y_b(10890)$. Therefore, we conclude that this state could also be 
described as a $[bq][\bar{b}\bar{q}]$ tetraquark state, in a scalar-vector 
configuration.

Considering the strange quark mass in the current (\ref{yb}) and doing the same 
analysis as before, we obtain the results shown in Fig.~\ref{FigYbs} for the 
$[bs][\bar{b}\bar{s}]$ tetraquark state. The valid Borel window in this case is: 
$(0.10 \leq \tau \leq 0.15) \GeV^{-2}$. Varying the continuum threshold in the 
range: $\sqrt{s_0} = 11.60 \pm 0.10 \GeV$, and the parameters as 
indicated in Table \ref{Param} we get:
\beq
 M_{Y_{bs}} = (10.97\pm0.10)~\GeV ~.
\label{ybs}
\enq
In this case the differences in the masses obtained for the non-strange and 
strange exotic tetraquark states are larger than is the case of the $\Upsilon~f_0$
and $\Upsilon~\sigma$ molecular currents.
Therefore, comparing the results presented in Table \ref{upsilon} and in 
Eqs.~(\ref{ybq}) and (\ref{ybs}), we conclude that
the $Y_b(10890)$ is better described by a $[bq][\bar{b}\bar{q}]$ 
tetraquark current, as suggested in ref.~\cite{ali}.

\section{Conclusions}

We have studied the mass of the $1^{--}$ exotic states using QCD sum rules. 
We find that the molecular currents $J/\psi~f_0(980)$ and $J/\psi~\sigma(600)$ 
lead to (almost) the same mass predictions, the difference being only around 
30 MeV. The mass obtained for the molecular state is smaller
 when the vector and scalar mesons are in the color singlet configuration, 
as compared with the color octet configuration. The masses obtained with the
 $J/\psi~f_0(980)$ and $J/\psi~\sigma(600)$ currents are in agreement with the
$Y(4660)$ mass. This result is not compatible with the proposition presented in 
ref~\cite{oset}, that the $Y(4260)$ state could be a $J/\psi~f_0(980)$ molecular
state. Since the mass obtained with the $J/\psi~f_0(980)$ current is above the
$J/\psi~f_0(980)$ threshold, we conclude that this current could be describing
a $\psi(2S)~f_0(980)$ loosely bound molecular state, that could be associated with 
the $Y(4660)$, as suggested in \cite{guo}.

We have also studied the    $\Upsilon~f_0(980)$ and $\Upsilon~\sigma(600)$ 
 molecular currents, in color singlet and color octet configuration for the mesons.
Also in this case the masses obtained with the
 $\Upsilon~f_0$ and $\Upsilon~\sigma$ currents are in agreement with each other,
and the mass obtained for the molecular state is smaller
 when the two constituents mesons are in the color singlet configuration. 

Finally we have studied the scalar-vector tetraquark current in the strange and
no-strange sectors. We find that the newly observed $Y_b(10890)$ state
can be well described by a  $[bq][\bar{b}\bar{q}]$  tetraquark current, as 
suggested in ref.~\cite{ali}. However, considering the uncertainties, the
$\Upsilon~f_0$ and $\Upsilon~\sigma$ molecular currents cannot be discarded.
More experimental data on the $Y_b(10890)$ decay channels could be used to
discriminate between different  assignments.

\section*{Acknowledgements}
This work has been partly supported by FAPESP and CNPq-Brazil. We would like to 
thank Prof. S. Narison for bringing our attention to the work in ref.~\cite{ali}
and for discussions at the beginning of this calculation.

\appendix
\section{Spectral Densities}\label{App1}
The spectral densities expressions for $J/\psi \:f_0(980)$, $J/\psi \:\sigma(600)$,
$\Upsilon \:f_0(980)$ and $\Upsilon \:\sigma(600)$ molecular currents, were 
calculated up to 
dimension-6 condensates, in the $g_{\mu\nu}$ structure, at leading order in 
$\alpha_s$. We have
kept terms which are linear in the strange quark mass $m_s$.  To keep the heavy
quark mass finite, we use the momentum-space expression for the heavy quark 
propagator. We calculate the light quark part of the correlation
function in the coordinate-space, and we use the Schwinger parameters to 
evaluate the heavy quark part of the correlator. To evaluate the $d^4x$
integration in Eq.(\ref{2po}), we use again the Schwinger parameters, after 
a Wick rotation. Finally we get integrals in the Schwinger parameters that can 
be performed after scaling these parameters and introducing a delta function 
in the scale parameter \cite{itzy}. The result of these integrals are given in 
terms of logarithmic functions, from where we extract the spectral densities 
and the limits of the integration. The same technique can be used to evaluate
the condensate contributions. For that we have only to use the OPE expansion
for the propagators. For the light quark propagator we use \cite{Nielsen:2009uh}:
\beqa
&&S_{ab}(x)=\lag{0} T[q_a(x)\overline{q}_b(0)]\rag{0}={i\delta_{ab}\over2
\pi^2x^4}\xsla-{m_q\delta_{ab}\over4\pi^2x^2}
\nonumber\\
&-&{t^A_{ab}g G^A_{\mu\nu}\over32\pi^2}\left({i\over x^2}
(\xsla\sigma^{\mu\nu}+\sigma^{\mu\nu}\xsla)-{m_q}\sigma^{\mu\nu}\ln(-x^2)\right)
\nonumber\\
&-&{\delta_{ab}\over12}\qq
+{i\delta_{ab}\over48}m_q\qq\xsla-{x^2\delta_{ab}
\over2^6\times3}\mix
\nonumber\\
&+&{ix^2\delta_{ab}\over2^7\times3^2}m_q\mix\xsla,
\label{proleve}
\enqa
where we have used the fixed-point gauge. For heavy quarks, as explained above, 
we work in the momentum space and the OPE expansion for the heavy quark propagator
is given by: 
\beqa
S_{ab}(p)&=&
i \frac{\psla + m}{p^2 - m^2} \delta_{ab}+
\nonumber\\\
&-&{i\over4} \frac{t^A_{ab}g G^A_{\mu\nu} 
[\sigma^{\mu\nu} (\psla + m) + (\psla + m) \sigma^{\mu\nu}]}{(p^2 - m^2)^2}
\nonumber\\
&+&\frac{i \delta_{ab}}{12} m\Gd\frac{p^2 + m\psla}{(p^2 - m^2)^4}.
\label{propesado}
\enqa

We have considered the compact 
notation for both color singlet and octet configurations, since the difference 
between them is only proportional to a color factor. For this we define:
\[
{\footnotesize
\begin{array}{ccc}
&\\
\hline
\mbox{Color Factor} ~~&~~ \mbox{Mesons in} ~~&~~ \mbox{Mesons in}\\
&~~ \mbox{Color Singlet} ~~&~~ \mbox{Color Octet}\\
\hline
\mathcal{N} & 9 / 2^5 & 1 \\
\mathcal{N}^\ast & -9 / 2^2 & 1\\
\hline\\
\end{array}}\]
where we have normalized the factor to the color octet configuration. 
Therefore, we obtain the following expressions:
{\footnotesize
\begin{eqnarray*}
\label{rhoope}
\rho^{pert}_{_M}(s)&=&{\mathcal{N} \over 3\cdot 2^{7} \pi^6}\!\!
\int\limits_{\almi}^{\alma}\!\!\!\!{d\al\over\alpha^3} \!\!
\int\limits_{\bemi}^{1-\al}\!\!\!\!{d\be\over\be^3}(1-\al-\be) F_Q^3(\al,\be) \\
&& \times \left[ (1+\al+\be)F_Q(\al,\be) - 4m_Q^2(1-\al-\be) \right], \\
\rho^{\qqs}_{_M}(s)&=&{\mathcal{N} m_s\qqs \over 4 \pi^4}
 \Bigg[ \int\limits_{\almi}^{\alma}\!\!\!\! {d\al} {H_Q^2(\al) \over \al(1-\al)} - 
\!\!\int\limits_{\almi}^{\alma}\!\!\!\! {d\al \over \al} \!\!\int
\limits_{\bemi}^{1-\al}\!\!\!\!{d\be\over\be} \\
&& \times F_Q(\al,\be) \left( F_Q(\al,\be) + 2m_Q^2 \right) \Bigg], \\
\rho^{\lag G^2\rag}_{_M}(s) &=& -{\Gd\over 3^2 \cdot 2^{11}\pi^6}
\Bigg\{ 6 \mathcal{N}^\ast \!\!\int\limits_{\almi}^{\alma}\!\!\!\!{d\al}
\frac{H_Q^2(\al)}{\al(1-\al)} + 32\mathcal{N} m_Q^2 \\
&& \times \!\!\int\limits_{\almi}^{\alma}\!\!\!\!{d\al \over \al^3} \!\!
\int\limits_{\bemi}^{1-\al}\!\!\!\!{d\be\over\be} (1-\al-\be) \Bigg[ (1-\al-\be)\\
&& \times (3 F_Q(\al,\be) + m_Q^2\be) -\be(1\!+\!\al\!+\!\be)F_Q(\al,\be) \Bigg] \\
&&- \mathcal{N}^\ast \!\!\!\int\limits_{\almi}^{\alma}\!\!\!\!{d\al \over \al^2} 
\!\!\!\int\limits_{\bemi}^{1-\al}\!\!\!\!{d\be\over\be^2} F_Q(\al,\be) 
\Bigg[ (1\!-\!\al\!-\!\be)F_Q(\al,\be)\\
&& \times (3\!-\!\al\!-\!\be) + 6m_Q^2(2\!+\!\al\!+\!\be)\al\be -6\al^2\be^2 s 
\Bigg], \\
\rho^{\mixs}_{_M}(s)&=&-{m_s\mixs\over 3^2\cdot 2^{4} \pi^4} (32\mathcal{N} + 
3\mathcal{N}^\ast)
\!\!\int\limits_{\almi}^{\alma}\!\! {d\al} ~\al(1-\al) s, \\
\rho^{\qqs^2}_{_M}(s)&=&-{4\mathcal{N} \qqs^2 \over 9\pi^2}\!\!\int
\limits_{\almi}^{\alma}\!\! {d\al} ~\al(1-\al) s
\end{eqnarray*}}

The spectral densities expressions for $[bq][\bar{b}\bar{q}]$ and $[bs]
[\bar{b}\bar{s}]$ tetraquark states with the current defined in Eq.(\ref{yb})
are given by:
{\footnotesize
\begin{eqnarray*}
\rho^{pert}_{_T}(s)&=&-\frac{1}{3 \cdot 2^{10} \:\pi^6} 
	\int\limits^{\alpha_{max}}_{\alpha_{min}} \!\!\!\!\frac{d\alpha}{\alpha^3} 
\!\!\!\!
	\int\limits^{1-\alpha}_{\beta_{min}} \!\!\!\!\frac{d\beta}{\beta^3}
	(1-\al-\be) F_b^3(\al,\be) \\
	&& \times \Bigg[ 2 m_b^2 \left(1 - \al - \be \right)^2 -3(1+\al+\be) 
F_b(\al,\be) \Bigg] \\
\rho^{\qq}_{_T}(s)&=&\frac{m_q \qq}{2^{5} \:\pi^4} \Bigg\{
	\int\limits^{\alpha_{max}}_{\alpha_{min}} \!\!\!\!\frac{d\alpha}{\al(1-\al)} 
	\left[ m_b^2 - \al(1-\al)s \right]^2 \\
	&& \hspace{-0.8cm} -\!\! \int\limits^{\al_{max}}_{\al_{min}} \!\!\!\!
\frac{d\al}{\al} \!\!\!
	\int\limits^{1-\al}_{\be_{min}} \!\!\!\!\frac{d\beta}{\beta}
	F_b(\al,\be) \!\Bigg[ \!m_b^2(5 \!-\! \al \!-\! \be) + 2 F_b(\al,\be) \!
\Bigg] \!\Bigg\} \\
\rho^{\lag G^2\rag}_{_T}(s)&=&-\frac{\Gd}{3^2 \cdot 2^{11} \:\pi^6} \!\!\!\!
	\int\limits^{\alpha_{max}}_{\alpha_{min}} \!\!\!\!\frac{d\alpha}{\al} \!\!
	\int\limits^{1-\alpha}_{\beta_{min}} \!\!\!\!\frac{d\beta}{\beta^3} \Bigg\{
	2m_b^4 \al(1\!-\! \al \!-\! \be)^3 \\
	&& \!\!+ 3 m_b^2 (1\!-\! \al \!-\! \be) \!\Bigg[ \!2 \!-\! 2\al(4 \!+\! 
\al \!+\! \be) \!-\! \be(1 \!+\! \al \!+\! \be) \!\Bigg] \\
	&& \times F_b(\al,\be) + 6\be \left( 1-2\al-2\be \right) F_b^2(\al,\be) 
\Bigg\} \\
\rho^{\mix}_{_T}(s)&=&\!-\frac{\mix}{3\cdot 2^{7} \:\pi^4} \Bigg\{ \!\!
	2m_q \!\!\int\limits^{\alpha_{max}}_{\alpha_{min}} \!\!\!\!\frac{d\alpha}
{\al} 
	\Bigg[ \!8m_b^2 \al + (1\!-\! \al) H_b(\al) \!\Bigg] \\
	&&- 3 m_b \!\!\int\limits^{\alpha_{max}}_{\alpha_{min}} \!\!\!\!
\frac{d\alpha}{\al}\!\!
	\int\limits^{1-\al}_{\be_{min}} \!\!\!\!\frac{d\be}{\be^2} \!
\Bigg[ (1 \!-\! \al \!-\! \be) + 2\al(\al \!+\! \be) \!\Bigg] \\
	&&  \times F_b(\al,\be) - m_q \!\!\int\limits^{\alpha_{max}}_{\alpha_{min}} 
\!\!\!\!d\alpha \!\!
	\int\limits^{1-\al}_{\be_{min}} \!\!\!\!\frac{d\be}{\be} \\
	&& \times \Bigg[ m_b^2 (9 \!-\! 3\al \!-\! 5\be) + 7 F_b(\al,\be) \Bigg]  
\Bigg\} \\
\rho^{\qq^2}_{_T}(s)&=&-\frac{\qq^2}{3\cdot 2^{3} \:\pi^2} 
	\int\limits^{\alpha_{max}}_{\alpha_{min}} \!d\alpha
	\left[ 3m_b^2 - \al(1-\al)s \right].
\end{eqnarray*}
}

In all expressions we have used the following definitions:
\beqa
F_Q(\al,\be) &=& m_{Q}^{2}(\al+\be) - \al\be s , \\
H_Q(\al) &=& m_{Q}^{2}-\al(1-\al)s
\enqa
and the integration limits are given by:
\beqa
\almi&=&({1-\sqrt{1-4m_Q^2/s})/2} ~, \\
\alma&=&({1+\sqrt{1-4m_Q^2/s})/2} ~, \\ 
\bemi&=&{\al m_Q^2/( s\al-m_Q^2)} ~.
\enqa
The index $Q = c ~\mbox{or}~ b$ indicates the heavy quark content in the current. 
We have neglected the contribution of the dimension-six condensate 
$\langle g_s^3 G^3\rangle$, since it is assumed to be suppressed by the loop 
factor $1/16\pi^2$.


\end{document}